\documentclass[aps, pra, a4paper, showpacs, twocolumn, english, 10pt]{revtex4-1}
\usepackage{bbm, amsthm, bm,textcomp, nicefrac,geometry,ragged2e}
\usepackage[dvipsnames]{xcolor}
\usepackage{float}
\usepackage[bbgreekl]{mathbbol}
\usepackage{graphicx,epstopdf,color,verbatim,enumitem,ulem}
\usepackage{pbox,array}
\usepackage{mathrsfs}
\usepackage{amssymb}
\usepackage{textgreek}
\usepackage{mathtools}
\usepackage{stackrel}
\usepackage[thinlines]{easytable}
\usepackage{amsmath}
\usepackage[utf8]{inputenc}
\makeatletter
\usepackage{soul}
\usepackage[caption=false]{subfig}
\usepackage{hyperref}
\hypersetup{
    colorlinks = true,
    linkcolor = Purple,
    citecolor = Purple,
    filecolor = Purple,      
    urlcolor = Purple,
}
\usepackage[T1]{fontenc}
\usepackage[caption=false]{subfig}
\usepackage{babel}
\usepackage[linesnumbered,ruled]{algorithm2e}

\addto\captionsenglish{}

\newcommand\id{\mathbbm{1}}

\newcommand{\ket}[1]{\left|#1\right\rangle}

\newcommand{\bra}[1]{\left\langle #1\right|}

\newcommand{\proj}[1]{\ket{#1}\!\bra{#1}}

\definecolor{brickred}{rgb}{0.8, 0.0, 0.0}

\begin{document}

\title{Collective Operations Can Exponentially Enhance Quantum State Verification}

\author{Jorge Miguel-Ramiro}
\thanks{These two authors contributed equally to this work.} 
\author{Ferran Riera-S\`abat}
\thanks{These two authors contributed equally to this work.} 

\author{Wolfgang D\"ur}
\affiliation{Institut f\"ur Theoretische Physik, Universit\"at Innsbruck, Technikerstra{\ss}e 21a, 6020 Innsbruck, Austria}

\begin{abstract}
Maximally entangled states are a key resource in many quantum communication and computation tasks, and their certification is a crucial element to guarantee the desired functionality. We introduce collective strategies for the efficient, local verification of ensembles of Bell pairs that make use of an initial information and noise transfer to few copies prior to their measurement. In this way the number of entangled pairs that need to be measured and hence destroyed is significantly reduced as compared to previous, even optimal, approaches that operate on individual copies. Moreover the remaining states are directly certified. We show that our tools can be extended to other problems and larger classes of multipartite states.
\end{abstract}

\maketitle

\textit{Introduction.---}With the emergence of quantum technologies, the certification and verification of quantum devices and states have become necessary requirements for viable quantum communication and computation tasks, such as e.g. quantum teleportation \cite{Bennett_telep}, quantum key distribution \cite{Ekert91,Bennett2014}, and distributed or blind quantum computation \cite{CiracDistributed,Hayashi15,Gheorghiu2019}. In particular, certification of maximally entangled states by local operations is a crucial ingredient for a feasible implementation of bottom-up \cite{Kimble2008,Wehner2018,Kozlowski2019,Azuma2021} and entanglement-based \cite{Pirker2018,Pirker2019,Meignant2019,Gyongyosi2019} quantum networks, where entanglement is a key resource to enable e.g. long-distance communication, various security applications or connecting distributed quantum processors. However, local measurements destroy entanglement, making the verification of entangled states costly. 

Different approaches for certifying quantum states exist \cite{Eisert2020,Yu2021}. Some of them, as state tomography \cite{Tomography2010}, are, however, very inefficient as all elements of the density matrix need to be determined by means of destructive measurements. A protocol called quantum state verification was introduced in \cite{Pallister18}, allowing for efficient verification of quantum states with local measurements and constant overhead with regard to optimal global strategies. Several extensions \cite{Li19,Wang2019,Bdescu2019,Zhu2019,Morris21} have been proposed, and were implemented experimentally \cite{Zhang2020}. These approaches rely in general on suitable sequential pass-or-fail measurements applied on individual states. However, the improved control of quantum systems also makes feasible more advanced, collective strategies that operate jointly on multiple copies. 

Here we show that such a collective but local strategy can significantly improve the efficiency of previous, even global and optimal, strategies based on sequential measurements of single copies. Our approach operates on multiple copies of entangled states, where only a few of these states are designated for certifying the whole ensemble. This is achieved by accumulating the noise of the whole ensemble into a reduced set of states by collective local operations, so that by measuring and consuming only these states one can detect the noise with enhanced probability while certifying the remaining states without destroying them. This significantly reduces the amount of entanglement that is destroyed due to the certification process. We adapt techniques from entanglement purification \cite{Riera1,Riera2} in order to transfer noise from states in the ensemble into a few target states that are then measured. Crucially, the nonmeasured states remain untouched and hence entangled, and can still be used as a resource for various nonlocal quantum tasks. Although we focus on maximally entangled Bell states throughout this Letter, we remark that our techniques can be extended to different quantum states, including, e.g., maximally entangled qudit states or multipartite Greenberger-Horne-Zeilinger states. 

\textit{Problem statement.---} Consider an ensemble of $n$ copies of some bipartite entangled state $\rho_{AB}$ shared by two parties $A$ and $B$, ideally prepared in the maximally entangled state $\proj{\Psi_{00}}$, where $\ket{ \Psi_{ij} }_{AB} = \id \otimes \sigma^j_x \sigma^i_z ( \ket{ 00 }_{AB} + \ket{ 11 }_{AB} ) / \sqrt{2}$ are the four Bell states. There is the promise \cite{Pallister18} that the states are all either perfect, i.e., $ \rho_{\Psi_{00}} = \proj{\Psi_{00}}$, or they have some noise corresponding to a mixed state $\rho$ with unknown fidelity $F = \left\langle \Psi_{00}\right|\rho\left|\Psi_{00}\right\rangle \leq 1 - \epsilon$. Some verification device is able to perform local operations on the parts of the states at $A$ and $B$ with the task of discerning which is the case, up to some failure probability $\delta_{\text{fail}}$. In this process, part of the ensemble is destroyed in order to examine whether $F = 1$. If that is the case, the conclusion is extended to the whole ensemble. Otherwise, all states are discarded. We show how our collective approach outperforms previous optimal strategies based on individual measurements \cite{Pallister18,Li19,Wang2019,Bdescu2019,Zhu2019,Morris21}.

\textit{The counter gate and d-level systems.---} Our protocol relies on a $d$-level auxiliary bipartite entangled state used to encode information of the whole ensemble. In particular, we denote the $d$-dimensional maximally entangled states as $\ket{\Phi^d_{mn}}_{AB} = \sum_{ k = 0 }^{ d - 1 } e^{ \text{i} 2\pi km /d } |k\rangle_A |k \ominus n \rangle_B / \sqrt{d}$, where $k \ominus n \equiv (k-n) \text{mod}\, d $, and the index $n$($m$) is called the amplitude (phase) index. The auxiliary state is used to accumulate and measure the noise of an ensemble of multiple noisy states. This is achieved by means of the so-called ``counter gate'' \cite{Riera1,Riera2} that transfers information from the ensemble of qubit states into the amplitude index of the auxiliary. This amplitude index can be read by locally measuring the state in the computational basis. The counter gate is defined as a bilateral controlled-$X$ gate, acting from a qubit pair as source, to a qudit pair as target. Notice that the gate can be implemented locally. If the target system is in a maximally entangled state with phase index zero, its action is given by
\begin{equation}
    \label{eq:acction:CX}
    b\text{CX}^{AB}_{ 1 \to 2} \! \ket{mn}_{\!A_1 B_1} \! \ket{ \Phi_{ 0j }^d }_{\!A_2 B_2} \!\! = \! \ket{mn}_{\!A_1 B_1} \! \ket{ \Phi^d_{ 0, j\ominus m \oplus n } }_{\!A_2B_2} \!,
\end{equation}
where $b \text{CX}^{AB}_{1 \to 2} = \text{CX}_{A_1 \rightarrow A_2} \otimes \text{CX}_{B_1 \rightarrow B_2},$ and $\text{CX}_{1\rightarrow 2}$ is the hybrid \textit{controlled-X} gate \cite{Daboul2003} $\text{CX}_{1 \rightarrow 2} = \proj{0} \otimes \id_d + \proj{1} \otimes X_d$. For convenience, we denote as type-1, type-2 and type-3 error states, the states corresponding to $|01\rangle$, $|10\rangle$ and $\ket{\Psi_{10}}$ respectively. The action of the counter gate, Eq. \eqref{eq:acction:CX}, with a type-1 (type-2) error state acting as control, leads to an amplitude index value of the auxiliary state increased(decreased) by 1, whereas it is left invariant if the control is a type-3 error state. Importantly, this invariance property also applies in cases in which the control system is in the $\ket{\Psi_{00}}$ state.

\textit{Proof of concept.---} We provide a basic example based on simplified assumptions in order to illustrate the details of our procedure, the so-called ``general error number gate protocol''. One can, however, relax these assumptions to tackle a completely general situation (see below).

Consider an ensemble of $n$ copies with the promise that all the states are either perfect Bell states $\proj{\Psi_{00}}$, or rank-2 states with only type-1 errors, i.e., $\rho = F \proj{\Psi_{00}} + (1-F)\proj{01}$. This corresponds (up to local unitaries) to a situation where independent decay channels act on a maximally entangled state $\proj{\Psi_{10}}$. Physically this relates to the decay of electronic excitations in atomic or ensemble-based quantum memories, but also to photon loss of photon-number states.

\begin{figure}
    \centering
    \includegraphics[width=\columnwidth]{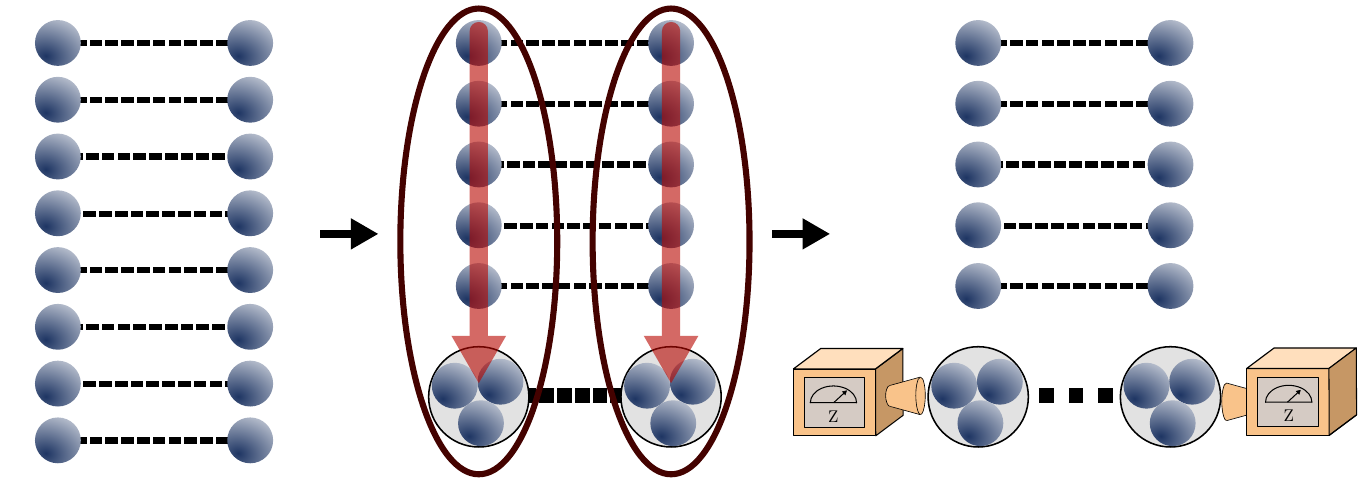}
    \caption{\label{fig:1} Schematic representation of the protocol. The ENG is applied from the ensemble states to concentrate the noise into an auxiliary qudit state (which can be constructed by embedding ensemble copies). Finally, the $d$-level system is locally measured to detect the noise while the rest of copies are left untouched.}
\end{figure}

The protocol comprises the following steps (see Fig. \ref{fig:1}). First, we apply the counter gate, Eq. \eqref{eq:acction:CX}, from each state in the ensemble to an auxiliary pure state $\proj{\Phi_{00}^d}$ with $d=n+1$. We show below that the auxiliary state can be constructed directly from the--noisy--ensemble copies. We denote these local operations together as the error number gate (ENG). The ENG changes the amplitude index of the auxiliary state depending on the actual form of the ensemble: $(i)$ Pure ensemble: the ensemble is given by $n$ copies of the $\proj{\Psi_{00}}$ state, and the application of the ENG leaves the auxiliary state invariant. $(ii)$ Noisy ensemble: the ensemble is given by $\rho^{\otimes n}$, and it can hence contain type-1 error states. Whenever the counter gate is applied with a single type-1 error state, the amplitude index of the auxiliary state is increased by 1. After the application of the ENG, the ensemble and auxiliary states get correlated, i.e., $\text{ENG:} \; \rho^{\otimes n} \otimes \proj{\Phi_{00}^d} \to \sum_{j=0}^n \binom{n}{j} F^{n-j} (1-F)^j \, \Gamma_j \otimes \proj{\Phi_{0j}^d}$, where $\Gamma_j$ is a density operator corresponding to all permutations of $\big\{ \ket{\Psi_{00}}^{ \otimes (n-j) } \ket{01}^{\otimes j} \big\}$. By measuring the auxiliary state, we learn the value of $j$, each found with probability $p(j) = \binom{n}{j}F^{n-j}(1-F)^j$, that depends on the state fidelity $F$. In this case, the value of $j$ indeed corresponds to the actual number of errors in the ensemble.

Whenever a value $j \neq 0$ is found, we can assert with certainty that we are in case $(ii)$ and the ensemble is noisy with $F<1$. On the other hand, if we obtain $j = 0$ we conclude, with some success probability, that the states of the ensemble are perfect Bell pairs $F = 1$ [case $(i)$]. In particular, the failure probability, i.e., measuring $j=0$ while the initial state was $\rho^{\otimes n}$, is $\delta_{\text{f}} = F^n$. In this case we failed to identify the noisy ensemble, and would draw a wrong conclusion. For a fixed failure probability, one can determine the minimum number of ensemble states $n$ (and therefore the minimum dimension of the auxiliary state) necessary to identify the case $(i)$. Notice that the dimension $d$ of the auxiliary state increases linearly with $n$, leading to an amount of entanglement (ebits) that only scales logarithmically with $n$, $\mathcal{O} \left( \log n \right)$. As we show below, this corresponds to the number of states from the initial ensemble that needs to be measured and destroyed.

A further improvement is possible. Since we are only interested in detecting whether $j\not=0$, directly measuring the whole auxiliary state might not be the most efficient strategy. By performing a two-outcome measurement on each part of the auxiliary state, of the form $\left\{ P_1, \id - P_1 \right\}$, where $P_1 = \sum_{i = 0}^{d/2 - 1}\proj{2i}$ (note the sum up to $d/2-1$), one can determine the parity of $j$. Same(different) outcomes in $A$ and $B$ correspond to an even(odd) $j$ value. We denote this protocol as the ``ENG subspaces protocol''. Whenever $j$ odd is obtained, we know with certainty that $j \neq 0$, and the ensemble is noisy [case $(ii)$]. In this case, one can recover the entanglement by performing an uncorrelating operation between the auxiliary state and the ensemble we leave the auxiliary system in the $\big|\Phi_{0,j/2}^{d/2}\big\rangle\big\langle\Phi_{0,j/2}^{d/2}\big|$ state (see Appendix~\ref{appendixA} for details). The case $(ii)$ is identified by consuming only 1 ebit. On the other hand, if $j$ even is found, the ensemble is considered to be perfect [case ($i$)] up to some failure probability $\delta_1$, which is now given by the probability of measuring that $j$ is even, while the ensemble is still noisy, $\delta_1 = \sum_{j=0}^{n/2} \binom{n}{2j} F^{n-2j} (1-F)^{2j}$. One can reduce $\delta_1$ by iteratively performing additional two-outcome measurements of the same form, learning--and consuming--1 ebit of information from the auxiliary state. The $m$th measurement can be written as $\{ P_m, \id - P_m \}$, where now
\begin{equation}
    P_m = \sum_{i = 0}^{\frac{d}{2m} - 1} \sum_{j = 0}^{m-1} \proj{2 m i \oplus j},
\end{equation}
revealing whether the value of $j$ is multiple of $2^m$ (or $0$). The failure probability, i.e., the probability of the ensemble being noisy and the outcomes of all $k$ measurements coinciding for $A$ and $B$ is
\begin{equation}
    \delta_m = \sum_{k=0}^{ n2^{-m} } \binom{n}{2^m k}F^{n-2^{m} k}(1-F)^{2^{m} k}.
\end{equation}
For some fixed $\delta_t$ one can obtain the number $(m)$ of measurements--number of ebits--required as a function of the ensemble fidelity $F$.

Observe that if an asymptotically large ensemble $n\to \infty$ is available, the required auxiliary entanglement needed for fixed $\delta_t$ becomes constant and independent of the fidelity of the initial states. In particular, the failure probability in the asymptotic case is $\delta_m = 2^{-m}$. The entanglement of the remaining subspaces is not spent or destroyed.

\textit{General case and results.---} We show here that all the assumptions can be relaxed and a completely general scenario can be tackled, exhibiting a performance enhancement with respect to previous approaches. We consider arbitrary ensembles, where importantly, the auxiliary state can be directly constructed from several copies of the ensemble.

We have the promise that all the ensemble states are either perfect Bell states or Werner states \cite{Werner89} of the form
\begin{equation}
    \label{eq:werner}
    \rho = q \proj{\Phi_{00}^d} + \frac{1 - q}{ d^2 } \, \id_{d^2},   
\end{equation}
with $d = 2$, where the fidelity is given by $F = (1 + 3q)/4$. This situation is completely general since any state can be brought to this form by applying random local operations \cite{Bennnett_1996}, without changing the fidelity. The protocol comprises the same steps as before, assuming for the moment (see below) that a maximally entangled state is available as auxiliary. However, one has to consider that now there are different kinds of errors, i.e., type-1 that increase, type-2 that decrease and type-3 that leave invariant the amplitude bit $j$ of the auxiliary state under the action of the ENG operation. A single copy of a Werner state can be interpreted as mixture of type-$1,2,3$ error states with probability $p_{1,2,3} = (1-F)/3$, and a Bell state with $p_0 = F$. Therefore, when applying the ENG from an ensemble of $n$ copies, the  value of the auxiliary amplitude index becomes $j = \Delta_{12} \, \text{mod}\, d$, where $\Delta_{12} = \#\text{type-1} - \#\text{type-2}$. The probability of obtaining a certain $j$ is given by
\begin{equation}
    \text{Pr}(j) = \! \sum_{\substack{i, k, \ell = 0 \\ i + k + \ell = n \\ k \ominus \ell = j } }^n \frac{n!}{i! \, k! \, \ell!} \, \left(p_0 + p_3 \right)^i p_1^k \, p_2^\ell.
\end{equation}
In each term of the sum, the number of type-1(type-2) errors is given by $k$($\ell$), and the number of states that are either $\ket{\Psi_{00}}$ or type-3 error state by $i$.

Note that the difference of errors can take $2n+1$ different values $\Delta_{12} \in \{-n, \dots, n\}$, and one would need an auxiliary state of $d=2n+1$ to distinguish between all of them. However, for our purpose we just need to determine when $\Delta_{12} = 0$, and therefore an auxiliary state of $d = n+1$ is sufficient, as $\Delta_{12} = 0 \Leftrightarrow \Delta_{12} \, \text{mod}(n+1) = 0$. The failure probability reads now $\delta = \text{Pr}_{(j = 0)}$.

\begin{table}
{\LinesNumberedHidden
    \begin{algorithm}[H]
        \SetKwInOut{Input}{Input}
        \SetKwInOut{Output}{Output}
        \SetAlgorithmName{Algorithm}{}
 \justifying \textit{Input}: Ensemble of $n$ identical quantum states, either $\ket{\Psi_{00}}$ or Werner-type states Eq.~\eqref{eq:werner} with $F<1$.
        \begin{enumerate}
            \item Construct an auxiliary state of $d = n + 1$ by embedding $\lceil \log_2 (n+1) \rceil$ ensemble states.
            \item Apply the ENG between the states of the ensemble and the auxiliary state.
            \item Locally measure the auxiliary amplitude index $j$.
        \end{enumerate}
\justifying  \textit{Output}: Information of noise of the ensemble. If $j \neq 0$ the noisy case is identified with $P=1$. If $j=0$, the ensemble is certified with $P=1-\delta$.
\caption{General ENG protocol overview}
\end{algorithm}}
\label{table:case1.Oneerror.}
\end{table}

We also consider here the subspaces ENG protocol. After measuring $m$ subspaces, and following the same steps as before, one obtains information about the $2^m$ multiplicity of the auxiliary amplitude index. In this case, the probability of failing in determining the noiseless scenario after measuring $m$ different subspaces is
\begin{equation}
    \delta_m = \sum_{k=0}^{\lfloor n 2^{-m} \rfloor} \text{Pr}( 2^m k ).
\end{equation}
Importantly, in the asymptotic limit we recover the constant , i.e., the number of copies for a fixed failure probability is insensitive to the fidelity of the initial states, such that $\delta_m = 2^{-m}$.

\begin{table}
{\LinesNumberedHidden
    \begin{algorithm}[H]
        \SetKwInOut{Input}{Input}
        \SetKwInOut{Output}{Output}
        \SetAlgorithmName{Algorithm}{}

       \justifying \textit{Input}: Ensemble of $n$ identical quantum states, either $\ket{\Psi_{00}}$ or Eq.~\eqref{eq:werner}.
        
        \begin{enumerate}
            \item Proceed as in Algorithm 1 steps 1-2.
            \item Parties $A$ and $B$ measure the subspace corresponding to the first Bell pair of the auxiliary. 
            \item If different outcome is found in $A$ and $B$, stop.
            \item Measure the next subspace until different outcome is found or enough $P_{\text{fail}}$ is achieved. 
        \end{enumerate}
        
    \justifying \textit{Output}: $2^k$ multiplicity of the value of $j$ after $k$ rounds. The noisy case is identified with certainty if measurement outcomes differ at any point, otherwise the ensemble is certified with $P=1-\delta$.
\caption{Subspaces ENG protocol overview}
\end{algorithm}}
\label{table:case1.Oneerrorsubpaces.}
\end{table}

So far we have assumed, for illustrative purposes, that a maximally entangled auxiliary state is available. This assumption is, however, not necessary, since the $d$-level auxiliary state can be always obtained by directly embedding--noisy--copies of the initial ensemble. Since the protocol is based on accumulating noise into the auxiliary state, by embedding several copies of the ensemble the performance is indeed enhanced, because noise already accumulates via embedding, before any other operation is applied. We define the embedding for perfect Bell states as $\big| \Phi_{00}^{2^k} \big\rangle_{\!\!AB} \!\! = \big|\Psi_{00} \big\rangle_{\!AB}^{\otimes k} = \sum_{i_1,\dots,i_k} \!\! \ket{i_k \dots i_1 }_{\!A} \ket{ i_k \dots i_1 }_{\!B}/\sqrt{2^k}$. This process with $m$ copies of noisy Bell states $\rho$ with fidelity $F$, leads to a noisy $d$-level state of $d = 2^m$. The resulting state can be always depolarized into an isotropic form \cite{Horodecki1999} of the form Eq.~\eqref{eq:werner}, with $d = 2^m$ and $q = (d^2 F^m - 1)/(d^2 - 1)$. If one directly measures the amplitude bit ($j$) of this state, before applying the ENG operation, the probability of measuring $j=0$ is given by $\delta = (1 + dF^m)/(1+d)$. The performance already approaches the optimal possible strategy based on measurements (see Appendix~\ref{appendixB} for details). The number of copies needed in this global optimal strategy based on single-copy measurements scales as $k = \ln{\delta} / \ln{F}$ \cite{Pallister18,Zhu19}.

One can, however, enhance the protocol performance--overcoming previous optimal single-copy strategies--by applying the ENG operation from the ensemble states into the auxiliary one. This process collects the noise of the ensemble into the auxiliary and, together with the noise already accumulated by the embedding, increases the probability of detecting the noise. As before, in case noise is detected, we discard all the ensemble, whereas if the noiseless case is identified, the ensemble is kept and certified, and only the auxiliary states are consumed.

In order to construct an auxiliary state of dimension $d = n+1$, which allows us to accumulate information about the noise of $n$ ensemble states, one just needs to embed $m = \log_2 (n + 1)$. Therefore, only $m$ copies are eventually consumed, since the dimension of the auxiliary scales exponentially with the number of embedded states, leading to an exponential improvement in the scaling and allowing us to overcome previous optimal bounds.

\begin{figure}
    \subfloat[\centering]{\includegraphics[width=0.5\columnwidth]{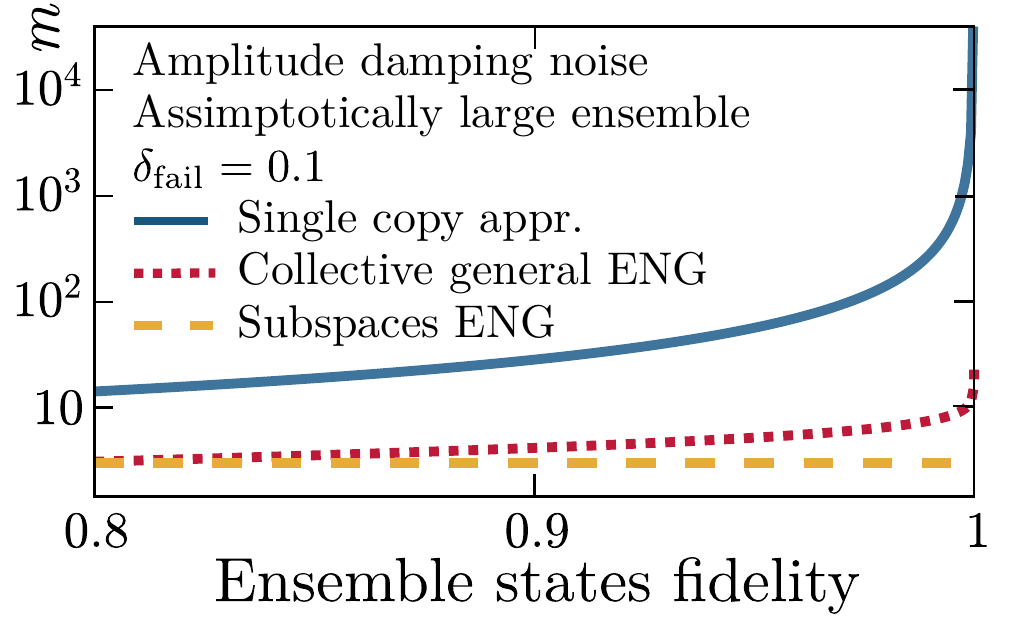}}
    \subfloat[\centering]{\includegraphics[width=0.5\columnwidth]{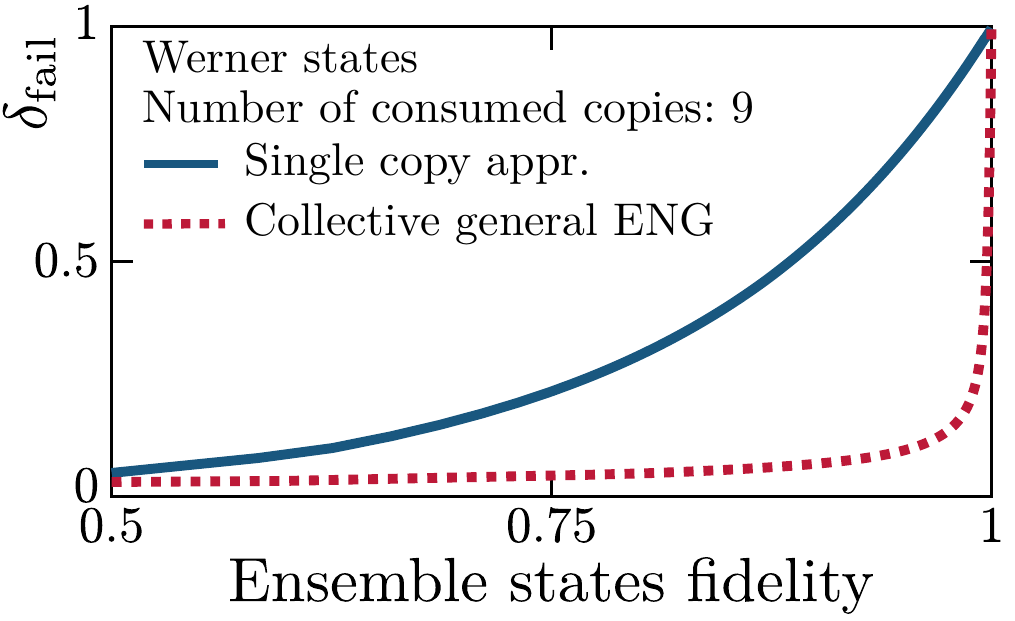}}
    \caption{\label{fig:2} Performance of the collective general ENG and the subspaces ENG protocols in comparison to optimal single-copy approach \cite{Pallister18}. In (a) the number of consumed copies for rank-2 states for a fixed failure probability $\delta_{\text{fail}} = 0.1$. In (b) the failure probability when $9$ copies are consumed for each strategy.}
\end{figure}

Figure \ref{fig:2} shows several results comparing the performance of our protocol with respect to the optimal approaches based on individual measurements, under different situations. One can see an exponential-type improvement in all the cases. In particular, if an arbitrarily large ensemble is available, the subspaces ENG strategy exhibits a constant behavior independent of the fidelity of the initial states. See Appendix~\ref{appendixC} for further analysis.

\textit{Generalizations.---} We have considered the verification of Bell states. However, the applicability of our approach goes beyond such states. In particular, these techniques can be applied to verify any set of states for which there exists a subspace that is invariant under the counter operations Eq.~\eqref{eq:acction:CX} (or equivalent). 

Some instances of states that can be verified include maximally entangled qudit states, or more general multipartite states. For the former case, the generalization is straightforward. Applying a generalized qudit-qudit controlled-$X$ \cite{Alber2001}, $G\text{CX} \ket{m} \ket{n} = \ket{m} \ket{n \oplus m}$, in a bilateral way between a bipartite qudit and a maximally entangled system of dimension $D$ \cite{dlevelpurif}, the effect is $bG\text{CX}\ket{m^d \, n^d}\ket{\Phi^D_{0j}} = \ket{m^d n^d}\ket{\Phi^D_{0,j \ominus n \oplus m}}$, where $j \ominus n \oplus n = (j - n + m) \text{mod } D$, similar than in the qubit case. Note that the dimension of the auxiliary should be adapted to the fact that errors can now increase or decrease the auxiliary amplitude bit by more than 1.

In a similar way, these techniques can be adapted to verify multipartite states. The invariant subspace of the generalized counter gate $m\text{CX}$ \cite{Riera2} is spanned by $\ket{00 \dots 0}$ and $\ket{11 \dots 1}$, while the amplitude vector of the auxiliary $d$-level system is modified depending on the error state. Therefore, a verification procedure for the Greenberger-Horne-Zeilinger state $(\ket{00\cdots0}+\ket{11\cdots1})/\sqrt{2}$ can be designed by extending the protocol for Bell states, since after applying the extended ENG the probability of obtaining a zero-valued amplitude index approaches zero when the number of copies in the ensemble increases. However, to make the procedure fully general extra operations are required to detect phase errors; see Appendix~\ref{appendixD} for details. Extension to more general graph states, following \cite{Dur2003,Kruszynska06}, might also be possible.

\textit{Conclusions.---} We have proposed collective techniques that allow us to verify maximally entangled quantum states with enhanced performance as compared to previous (even optimal) strategies that operate on individual states. This is accomplished by transferring and accumulating (via a so-called ENG operation) the noise of some ensemble of states into a higher-dimensional auxiliary state. This auxiliary state can be constructed using a logarithmically reduced number of ensemble copies, which are the only ones eventually consumed. Because of the embedding process and the ENG operation, noise is enlarged into the auxiliary state, making its detection more efficient. In addition, we propose a strategy based on measuring only certain subspaces of the auxiliary state, such that in the asymptotic limit of a large enough ensemble, a constant number of consumed copies is enough for verifying the states, independently on the fidelity or form of the states. The tools we introduce and make use of here are not only interesting in the context of certification of quantum states, but they can be particularly useful in other scenarios such as e.g. fidelity estimation (see Appendix~\ref{appendixE}) or fidelity witnessing (see follow-up work \cite{paperwit}). For the rank-2 example originating from decay noise, we can actually use our strategy not only to verify the ensemble, but to accurately estimate the fidelity by using only a logarithmic amount of extra entanglement, exponentially outperforming single-copy strategies.

\renewcommand{\acknowledgments}{}

\begin{acknowledgments}
This work was supported by the Austrian Science Fund (FWF) through
projects No. P30937-N27, No. P36009-N and No. P36010-N.
\end{acknowledgments}

\bibliographystyle{apsrev4-1}
\bibliography{Fidwitn_biblio}

\onecolumngrid
\renewcommand\appendixname{Appendix}
\appendix

\section{Uncorrelating the auxiliary state}
\label{appendixA}

In the main text, we show how one can verify Bell states by using qudits systems obtained from the same --noisy-- ensemble. However, one can also consider the approach where perfect auxiliary maximally entangled qudit states are used. This situation can be particularly useful in possible extensions of the protocol to perform different tasks, such as fidelity estimation. In this section, we show how by using pure auxiliary states, the protocol can be applied without consuming all the auxiliary entanglement.

Consider the Werner state with fidelity $F$, i.e., 
\begin{equation}
    \rho = q \, |\Psi_{00}\rangle \langle \Psi_{00} | + \frac{ 1-q }{4}\id_4,
\end{equation}
where $q = (4F-1)/3$. We rewrite the state as
\begin{equation}
    \label{eq:werner2}
    \rho = (p_0 + p_3) \, \sigma + p_1 \proj{01} + p_2 \proj{10}
\end{equation}
where $p_{1,2,3} = (1-F)/3$, $p_0 = F$ and
\begin{equation}
    \sigma = \frac{1}{p_0 + p_3} \big( p_0 \proj{\Psi_{10}} + p_3 \proj{\Psi_{10}} \big).
\end{equation}
describes a state that that can be either the type-3 error state or the target Bell state.

If we apply the ENG between and ensemble of $n$ copies of a Werner state, Eq.~\eqref{eq:werner2}, and an auxiliary maximally entangled state of the form 
\begin{equation}
    \ket{\Phi_{00}^{2^k}}_{AB} = \sum_{i_1,\dots,i_k = 0}^1 \ket{i_k, \dots, i_1 }_{A_k\dots A_1}\ket{i_k, \dots, i_1 }_{B_k \dots B_1},
\end{equation}
the state of the whole system is transformed into
\begin{equation}
    \label{app:eq:omega}
    \text{ENG:} \;\; \rho^{\otimes n}\otimes \proj{\Phi_{00}^d} \; \to \; \Omega = \sum_{j=0}^{d-1} p(j) \, \Gamma_j \otimes \proj{\Phi_{0j}^d},
\end{equation}
where
\begin{equation}
    p(j) = \sum_{\substack{i, k, \ell = 0 \\ i + k + \ell = n \\ k \ominus \ell = j} }^n \frac{n!}{i! \, k! \, \ell!} \left( p_0 + p_3 \right)^i \, p_1^k \, p_2^\ell
\end{equation}
is the probability of measuring a difference of errors given by $j = (k - \ell) \text{mod}\, d$ and $\Gamma_j$ is the density matrix describing the ensemble if a certain value of $j$ is obtained, i.e., 
\begin{equation}
    \Gamma_j = \frac{1}{p(j)} \sum_{\substack{i, k, \ell = 0 \\ i + k + \ell = n \\ k \ominus \ell = j} }^n (p_0 + p_3)^i \, p_1^k \, p_2^\ell \;\; \Pi \left[ \sigma^{\otimes i} \otimes \proj{01}^{\otimes k} \otimes \proj{10}^{\otimes \ell} \right]
\end{equation}
where $\Pi$ denotes the sum over all $n!/ (i! \, k! \, \ell !)$ permutations of the states.

After applying the ENG, we can obtain the parity of $j$ by locally measuring qubits $A_1$ and $B_1$ in the $Z$ basis, i.e., the measurement given by 
\begin{equation}
    \mathcal{M} : \left\{ M_1 = \proj{00}, \, M_2 = \proj{01}, \, M_3 = \proj{10}, \, M_4 = \proj{11}  \right\}.
\end{equation}
If the outcomes of both measurements coincide, i.e., if we obtain $M_1$ or $M_4$, then $j$ is even, otherwise $j$ is odd. This measurement modify the auxiliary state, which is transformed as
\begin{equation}
    \begin{gathered}
        M_{1,4} \ket{\Phi_{0j}^d} = \ket{ \Phi_{0,j/2}^{d/2} } \\
        M_2 \ket{\Phi_{0j}^d} = \ket{ \Phi_{0,(j-1)/2}^{d/2} } \\
        M_3 \ket{\Phi_{0j}^d} = \ket{ \Phi_{0,(j+1)/2}^{d/2} } \\
    \end{gathered}
\end{equation}
where systems $A_1$ and $B_1$ are no longer considered. After determining the parity of $j$ the state of the state of the ensemble also changes and whole system is given by
\begin{equation}
\begin{gathered}
    M_{1,4} \, \Omega \, M_{1,4} = \sum_{j \text{ even}} \Gamma_{j} \otimes \proj{ \Phi_{0,j/2}^{d/2} } \\ 
    M_2 \, \Omega \, M_2 = \sum_{j \text{ odd}} \Gamma_j \otimes \proj{ \Phi_{0,(j-1)/2}^{d/2} } \\ 
    M_3 \, \Omega \, M_3 = \sum_{j \text{ odd}} \Gamma_j \otimes \proj{ \Phi_{0,(j+1)/2}^{d/2} }
\end{gathered}
\end{equation}
This step can be iterated obtaining the parity of the new amplitude index, and so on until $j$ is fully determined. However, our protocol aborts if a $j\neq 0$ is obtained. Therefore, if the auxiliary state is not fully measured and we already obtained that $j \neq 0$, we can uncorrelate the auxiliary system keeping the entanglement left.

Then, we can take a copies of an auxiliary pure Bell state $\ket{\Psi_{00}}$, and embed it in the remaining auxiliary state, what duplicating its dimension by two, i.e., 
\begin{equation}
\begin{gathered}
    \sum_{j \text{ even}} \Gamma_{j} \otimes \proj{ \Phi_{0,j/2}^{d/2} } \otimes \proj{\Psi_{00}} = \sum_{j \text{ even}} \Gamma_{j} \otimes \proj{ \Phi_{0j}^d } \\ 
    \sum_{j \text{ odd}} \Gamma_j \otimes \proj{ \Phi_{0,(j-1)/2}^{d/2} } \otimes \proj{\Psi_{00}} = \sum_{j \text{ odd}} \Gamma_j \otimes \proj{ \Phi_{0,j\ominus 1}^d } \\ 
    \sum_{j \text{ odd}} \Gamma_j \otimes \proj{ \Phi_{0,(j+1)/2}^{d/2} } \otimes \proj{\Psi_{00}} = \sum_{j \text{ odd}} \Gamma_j \otimes \proj{ \Phi_{0,j\oplus 1}^d },
\end{gathered}
\end{equation}
where we use that
\begin{equation}
\begin{aligned}
    \ket{\Phi^d_{0j}}_{A_1 B_1} \ket{\Psi_{00}}_{A_2 B_2} & = \frac{1}{\sqrt{2 d}} \sum_{k=0}^{d-1} \big( \ket{k, 0}_{A_1 A_2} \ket{ k \ominus j, 0}_{B_1 B_2} + \ket{k, 1}_{A_1 A_2} \ket{ k \ominus j, 1}_{B_1 B_2} \big) \\ 
    & = \frac{1}{\sqrt{2d}}  \sum_{k' = 0}^{2d -1} \ket{k'}_A \ket{k'\ominus 2j}_B = \ket{\Phi^{2d}_{0,2j}}_{AB}.
\end{aligned}
\end{equation}
Then by applying a certain correction operation depending on the measurement outcome, we obtain the same state before taking any measure $\Omega$, Eq.~\eqref{app:eq:omega}. Note that this procedure can be iterated to recover the state $\Omega$ if $2m$ subsystems of the auxiliary system $A_m \dots A_1 B_m \dots B_1$ have been measured. In this case, we need $m$ copies of the Bell state $\ket{\Psi_{00}}$.

Once the state $\Omega$ is recovered, we uncorrelate the auxiliary system form the ensemble by applying the inverse of the ENG i.e.,
\begin{equation}
    \text{ENG}^{\dagger}\!: \;\; \Gamma_j \otimes \proj{\Phi_{0j}^d} \to \Gamma_j \otimes \proj{\Phi_{00}^d}.
\end{equation}
In this way, we can obtain the $m$-multiplicity of $j$ by consuming $m$ ebits.

\section{Alternative expression for the failure probability}
\label{appendixB}

Here, we introduce an alternative expression to compute the failure probability, i.e., the probability $\delta$ of measuring $j = 0$ but still being in the non-perfect case. The failure probability is given by
\begin{equation}
    \delta = \sum_{i=0}^n \binom{n}{i} q^{n-i}(1-q)^i \Omega_{\left(n-i\right)},
\end{equation}
where 
\begin{equation}
    \Omega_{ \left( s \right) } = \sum_{j = 0}^{ s/2 } \left( \frac{1}{4} \right)^{2j} \left( \frac{1}{2} \right)^{s - 2j} \frac{s!}{j! j! (s - 2j)!}
\end{equation}
determines the number of situations where the value of $j$ is left invariant --there are the same number of increasing and decreasing errors for each value $s=n-i$--. 

We also consider the approach where only certain subspaces of the auxiliary state are measured. After measuring $m$ subspaces, and following the same steps as before, one obtains information about the $2^m$-multiplicity value of the amplitude index $j$. In this case, the probability of failing in determining the perfect scenario after measuring $m$ different $2$-level states is
\begin{equation}
    \delta = 2 \sum_{t=0}^{2^{-m}n} \sum_{i=0}^{n} \binom{n}{i} q^iv(1-q)^{n-i}v\Omega_{\left(n-i,m\right)},
\end{equation}
where 
\begin{equation}
   \Omega_{\left(s,m\right)} = \sum_{j=0}^{s/2-2^{m-1}t} \left(\frac{1}{4}\right)^{j} \left(\frac{1}{4}\right)^{j+2^{m}t} \left(\frac{1}{2}\right)^{s-2j-2^{m}t} \frac{s!}{j!\left(j+2^{m}t\right)!(s-2j-2^{m}t)!}
\end{equation}
determines now the number of cases where the net sum of different error types is $m$-multiple of $2$. Importantly, in the asymptotic limit we recover the constant behavior, i.e., the number of copies for a fixed failure probability is insensitive to the fidelity of the initial states, such that $\delta_{\mathrm{fail}} = 2^{-k}$, where $k$ is the number of subspaces measured.

\section{Additional protocol performance analysis}
\label{appendixC}

We complete here the protocols analyses provided in the main text by including extra illustrative information for different problem settings.

\begin{figure*}
\centering
    \includegraphics[width=\textwidth]{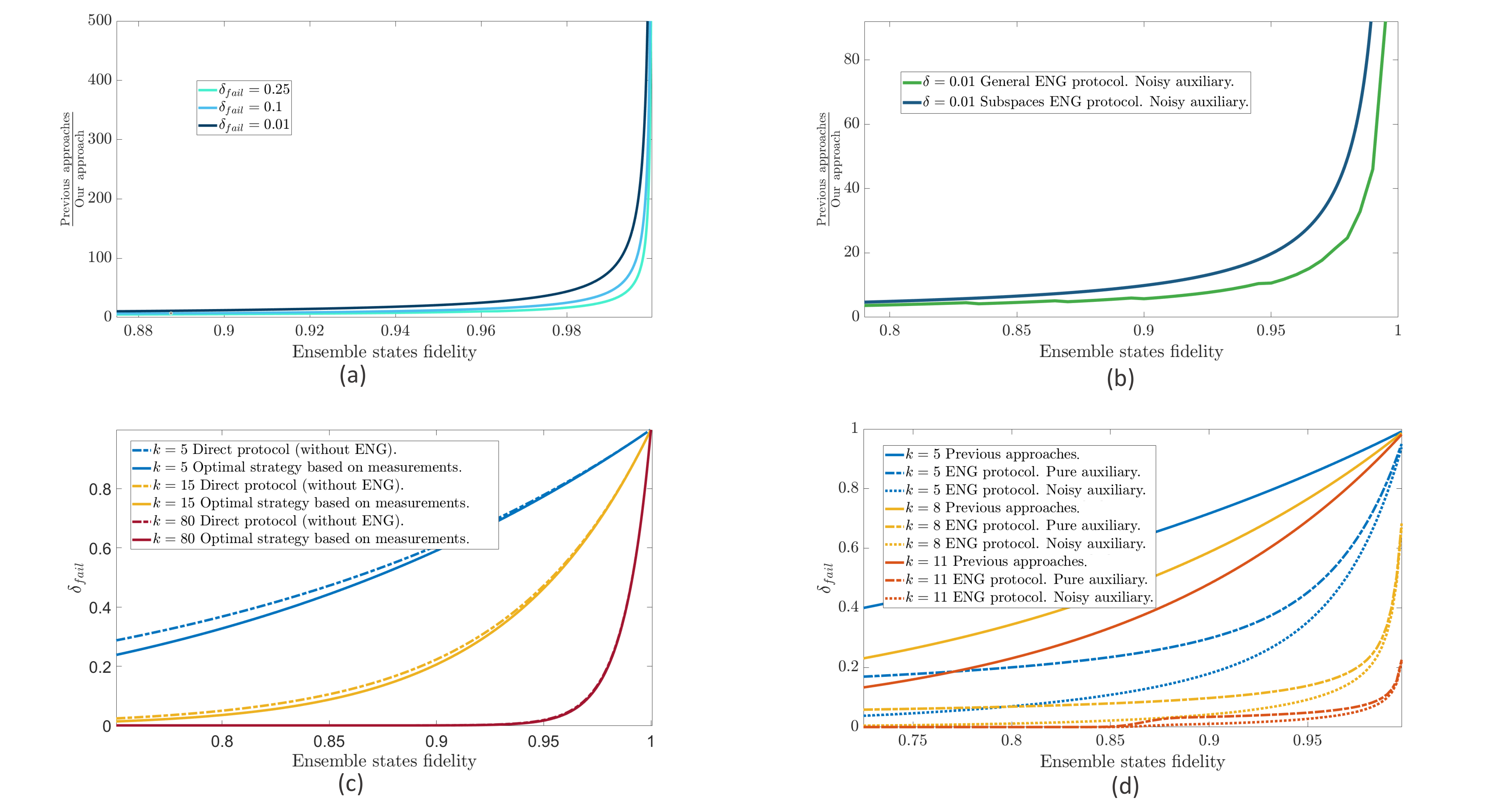}
    \caption{\label{fig:FIG1_Appendix} (a) Comparison for ensembles of rank-2 states of the ratio  between standard approaches and our approach performance.  (b) Performance ratio for Werner-type states for the general and the subspaces ENG protocols. (c) Comparison of the optimal global strategy and the case where the embedded state is directly measured (\textit{before} ENG).  (d) Failure probability for and initial ensemble of Werner states and different number of copies consumed. It is compared the approach assuming a pure auxiliary state and the approach consisting of building the --noisy-- auxiliary directly from the ensemble.  }
\end{figure*}

Fig.~\ref{fig:FIG1_Appendix} shows the performance of the protocols introduced in this work under different situations. Fig \ref{fig:FIG1_Appendix} (a), (b) represents the ratio or improvement of the protocols introduced in comparison to the best previously  known strategies, for rank-2 and general Werner states respectively. One can see an exponential improvement in both cases. In addition, Fig \ref{fig:FIG1_Appendix} (c) shows the advantages derived from using the noisy states of the ensemble to construct the auxiliary states. By doing so, and before the ENG that accumulate noise on the auxiliary, a measurement on the auxiliary system would already reveal information very close to the previous optimal strategies. Finally, Fig \ref{fig:FIG1_Appendix} (d) stresses the suitability of using copies of the ensemble to construct auxiliary systems, and the enhancement obtained with our approaches. 

\section{\textit{m}-party GHZ state}
\label{appendixD}

In a multipartite scenario, our verification protocol can be extended to verify $m$-partite GHZ states, i.e., the state given by
\begin{equation}
    \ket{\text{GHZ}_m}_{A B \dots M} = \frac{1}{\sqrt{2}} \, \Big( \ket{00 \dots 0}_{A B \dots M} + \ket{11 \dots 1}_{A B \dots M}\Big).
\end{equation}
Consider the orthonormal qubit GHZ-basis given by
\begin{equation}
    \ket{\Psi_{i\boldsymbol{j}}}_{A B \dots M} = \frac{1}{\sqrt{2}} \sum_{k=0}^1 e^{\text{i} \pi k i} \ket{k}_A \ket{k \oplus i_1}_B \cdots \ket{k \oplus i_{m-1}}_M,
\end{equation}
where $i$ is the phase bit and $\boldsymbol{j}$ is the amplitude bit vector. Note that our target state corresponds to $\ket{\text{GHZ}_m} = \ket{\Psi_{0\boldsymbol{0}}}$.
In a similar way as in the bipartite case, in this scenario the auxiliary system is given by a $d$-level $m$-partite GHZ state $\ket{\Phi^d_{0\boldsymbol{0}}}$, with
\begin{equation}
    \ket{\Phi^d_{i \boldsymbol{j}}} = \frac{1}{\sqrt{d}} \sum_{k = 0}^{d-1} e^{\text{i} \frac{2\pi}{d} k i} \ket{k} \ket{ k\ominus j_1} \cdots \ket{k \ominus n_{j-1}},
\end{equation}
where $i$ is the phase index and $\boldsymbol{j} = (j_1, \dots, j_{m-1})$ is the amplitude vector. We denote $k \ominus [\oplus] \, l \equiv (k - [+] \, l) \text{mod} \, d$. Note that, like in the bipartite case, one can obtain either the value of the phase index $i$ or the value of the amplitude vector $\boldsymbol{j}$ by measuring each qubit on the $X$ or on the $Z$ basis respectively and communicating the outcomes to the other parties afterwards.

We define the $m$-partite counter gate given by a multilateral control-$X$ gate with a qubit system as a control and a qudit system as target, i.e.,
\begin{equation}
    m\text{CX} = \text{CX}_{A_1 A_2} \otimes \text{CX}_{B_1 B_2} \otimes \cdots \text{CX}_{M_1 M_2},
\end{equation}
with $\text{CX}_{12} = \proj{0} \otimes \mathbbm{1} + \proj{1} \otimes X_d$.
The action of the counter gate with respect to a state of the computational basis acting as  control, and an a $d$-level system of the form $\big|\Phi^d_{0 \boldsymbol{j}}\big\rangle$ as target, is given by
\begin{equation}
\begin{aligned}
    m\text{CX} \ket{i_1}_{A_1} \cdots \ket{i_m}_{M_1} \ket{\Phi^d_{0 \boldsymbol{j}} }_{A_2\dots M_2} \\
    = \ket{i_1}_{A_1} \cdots \ket{i_m}_{M_1} \, \frac{1}{\sqrt{d}} \sum_{k=0}^{d-1} \ket{k \ominus i_1}_{A_2} \ket{k \ominus j_1 \ominus i_2}_{B_2} \cdots \ket{k \ominus j_{m-1} \ominus i_m}_{M_2} \\
    = \ket{i_1}_{A_1} \cdots \ket{i_m}_{M_1} \, \frac{1}{\sqrt{d}} \sum_{k'=0}^{d-1} \ket{k'}_{A_2} \ket{k' \ominus j_1 \ominus i_2 \oplus i_1}_{B_2} \cdots \ket{ k' \ominus j_{m-1} \ominus i_m \oplus i_1 }_{M_2} \\
    = \ket{i_1}_{A_1} \cdots \ket{i_m}_{M_1} \ket{\Phi^d_{0\boldsymbol{j}'}}_{A_2\dots M_2},
\end{aligned}
\end{equation}
where the components of $\boldsymbol{j}' = (j'_1, \dots, j'_{m-1})$ are given by
\begin{equation}
    j'_k = j_k \oplus i_{k+1} \ominus i_1.
\end{equation}
Note that the application of the counter gate transforms the amplitude vector of the auxiliary system for each state of the computational basis, except for the subspace $span\{ \ket{00 \dots 0} , \ket{11 \dots 1} \}$ that leaves the $d$-level system invariant. In other words, if the control system is of the form $\ket{\psi} =  \alpha \ket{00 \cdots 0} + \beta \ket{11 \cdots 1}$, the counter gate leaves the target system unchanged, i.e.,
\begin{equation}
    m\text{CX} \big( \alpha \ket{00 \cdots 0} + \beta \ket{11 \cdots 1} \big) \ket{\Phi^d_{0\boldsymbol{j}}} = \big( \alpha \ket{00 \cdots 0} + \beta \ket{11 \cdots 1} \big) \ket{\Phi^d_{0\boldsymbol{j}}}.
\end{equation}
Therefore, the target state $\ket{\Psi_{0\boldsymbol{0}}}$ keeps the auxiliary state invariant, and hence we can apply an analogous protocol as in the bipartite case, for noisy GHZ states. In the bipartite case, we can always depolarize the ensemble into an unknown collection of pure states, where each state is either the $\ket{\text{GHZ}_m}$ state or an error state. However, in the multipartite case, we find that some kinds of errors cannot be detected with by means of the counter gate.

As shown in \cite{wolfgang_ghz} one can always depolarize any noise GHZ state to the form
\begin{equation}
    \text{DEP:} \;\; \rho \to \varrho = F \proj{\Psi_{0\boldsymbol{0}}} + \lambda_0 \proj{\Psi_{1\boldsymbol{0}}} + \sum_{k=1}^{2^m-2} \lambda_k \proj{k},
\end{equation}
where $\ket{k} = \ket{k_m} \dots \ket{k_2} \ket{k_1}$, with $k_i$ being the $i$-digit of $k$ in the binary form, and $\ket{\bar{k}} = X_1 X_2 \cdots X_m \ket{k}$. In the depolarization procedure the fidelity $F$ of the state and the weight of $\proj{\Psi_{0\boldsymbol{0}}}$ given by $\lambda_0$ are kept, i.e.,
\begin{equation}
    \begin{gathered}
        F = \left\langle \Psi_{0\boldsymbol{0}} \right| \rho \left| \Psi_{0\boldsymbol{0}} \right\rangle = \left\langle \Psi_{0\boldsymbol{0}} \right| \varrho \left| \Psi_{0\boldsymbol{0}} \right\rangle \\
        \, \lambda_0 = \left\langle \Psi_{1\boldsymbol{0}} \right| \rho \left| \Psi_{1\boldsymbol{0}} \right\rangle = \left\langle \Psi_{1\boldsymbol{0}} \right| \varrho \left| \Psi_{1\boldsymbol{0}} \right\rangle.
    \end{gathered}
\end{equation}
In this case it is hence necessary to differentiate between two sources of errors: amplitude error and phase errors. Amplitude error are given by states of the computational basis orthogonal to the target system, i.e., $\{ \ket{k} \}_{k=0}^{d-2}$. On the other hand, phase errors are described by the state $\ket{\psi_{1\boldsymbol{0}}}$. Note that if we apply the counter gate with a amplitude error state acting as a control, the auxiliary state is modified and the error can be probabilistically detected. However, if the control system is in a phase error the auxiliary state remains invariant and it cannot be detected. In conclusion, our protocol can be used to verify unknown noisy GHZ state when they are affected by amplitude errors.

To verify general noisy GHZ states, we can proceed with the standard protocol to probabilistically detect amplitude errors. Then, in case of no errors detected we can apply a second round with a different control operation to detect the phase errors. For instance, with an auxiliary system of $d=2$, applying the counter gate with the auxiliary system as a control, i.e.,  
\begin{equation}
    m\text{CX}_{\text{aux}\to 1} \ket{\Psi_{m \boldsymbol{0}}}_1 \ket{\Psi_{n\boldsymbol{0}}}_{\text{aux}} = \ket{\Psi_{m\boldsymbol{0}}}_1 \ket{\Psi_{n\oplus m,\boldsymbol{0}}}_{\text{aux}},
\end{equation}
the phase bit of the auxiliary system is changed if the control system is a phase error and it is leaved invariant if the control system is the target state.

\section{Fidelity estimation for states resulting from decay}
\label{appendixE}

The tools and techniques introduced in this work can be applied for solving or improving other problems. In particular, we briefly discuss here how to efficiently tackle the fidelity estimation problem \cite{Flammia2011}. Further discussion and applications can be found in an upcoming work \cite{paperwit}. 

Given an ensemble of $n$ mixed entangled state, the task is to determine the fidelity of the states, up to some additive error  $F \pm \delta$ and failure probability $p_f$. The process entails the same steps as the ones for Algorithm 1 in the main text. Noise information of the ensemble is accumulated, via the local ENG operation, into a higher-dimensional auxiliary state, which is subsequently locally measured in order to learn the value of $j$, i.e. the number of errors contained in the ensemble.

The value of $j$ provides similar information as the one from previous strategies based of single-copy measurements \cite{Flammia2011}, but with exponentially reduced amount of resources spent. The probability of finding a certain value of $j$ for some fidelity $F$ is given by
\begin{equation}
    \text{Pr}(j|F) = \binom{n}{j} F^{n-j} (1-F)^j,
\end{equation}
and
\begin{equation}
    \label{eq:pj}
    \begin{aligned}
        \text{Pr}(j|F) = \sum_{\substack{i, k, l = 0 \\ i + k + l = n \\ k \ominus l = j } }^n \frac{n!}{i! \, k! \, l!} \, A^i \left( 1-A \right)^{k + l}
    \end{aligned}
\end{equation}
with $A \equiv (1+2F)/3$, for damping and depolarizing noise respectively, when an auxiliary maximally entangled state of dimension $d=n+1$ is available. In analogy to the protocols introduced in this work, similar expressions can be found by using directly the noisy states of the ensemble. To achieve the same accuracy as with a single-copy strategy \cite{Flammia2011} that measures $m$ noisy pairs directly, the required copies to construct the auxiliary state only scale as $\log(m)$. This provides an exponential enhancement. 

Given an ensemble of states with unknown fidelity $F$, and from the above probability distributions, one can evaluate the probability of determining that the fidelities lies in certain interval $F \pm \delta$. As in QSV, an important feature of this strategy together with the performance improvement, is the fact that the states whose fidelity is estimated are not destroyed in the process. 

\end{document}